\documentclass{cccg03}
\usepackage{amsfonts,url}

\title{Open Problems from CCCG 2002}
\author{%
  Erik D. Demaine%
    \thanks{MIT Laboratory for Computer Science,
            200 Technology Square, Cambridge, MA 02139, USA,
            \protect\url{edemaine@mit.edu}}
\and
  Joseph O'Rourke%
    \thanks{Department of Computer Science, Smith College, Northampton, MA
      01063, USA.
      \protect\url{orourke@cs.smith.edu}.
       Supported by NSF Distinguished Teaching Scholars award
       DUE-0123154.}}
\date{}

\let\latexcite=\cite
\def\cite{\nolinebreak\latexcite}
\let\latexref=\ref
\def\ref{\nolinebreak\latexref}

{\makeatletter
 \gdef\xxxmark{%
   \expandafter\ifx\csname @captype\endcsname\relax % based on a TeXBook e.g.
     \marginpar{xxx}% not in a caption, can use marginpar
   \else
     xxx % notice trailing space
   \fi}
 \gdef\xxx{\@ifnextchar[\xxx@lab\xxx@nolab}
 \long\gdef\xxx@lab[#1]#2{{\bf [\xxxmark #2 ---{\sc #1}]}}
 \long\gdef\xxx@nolab#1{{\bf [\xxxmark #1]}}
}

\def\captionfont{\sf}
\def\captionlabelfont{\bf}
{\makeatletter
 \global\let\plainfont@makecaption=\@makecaption
 \long\gdef\@makecaption#1#2{%
   \plainfont@makecaption{\captionlabelfont #1}{\captionfont #2}}}

\let\realsection=\section
\def\section*#1{\vspace{-2ex}\realsection*{\normalsize \bf #1}\vspace{-2ex}}

\newcommand\problem[4]
  {\item[\vtop{\hbox{\rlap{#1}}\hbox{#2}\hbox{#3}\hbox{#4}}]~\par
   \nopagebreak\medskip\nopagebreak}

\newbox\TOPPbox
\newcommand\TOPP[1]
  {\setbox\TOPPbox\vbox{\vspace{-4\baselineskip}\hbox to \hsize{\hfil
     \fbox{\vtop{\sf\settowidth\hsize{TOPP}%
                 TOPP\\\centerline{\##1}}}}}%
   \ht\TOPPbox=-1.2ex\dp\TOPPbox=0pt\box\TOPPbox}
\begin{document}
\maketitle

The following is a list of the problems presented on August 12, 2002 at
the open-problem session of the 14th Canadian Conference on Computational
Geometry held in Lethbridge, Alberta, Canada.

Boxed problem numbers indicate appearance in The Open Problem Project (TOPP);
see \url{http://www.cs.smith.edu/~orourke/TOPP/}.

\begin{description}

\problem{Great Circle Graphs: 3-colorable?}
        {Stan Wagon}{Macalester College}{wagon@macalester.edu}
\TOPP{44}

Is every zonohedron $3$-colorable when viewed as a planar map?
This question arose out of work described in~\cite{rsw-bo-02}.
An equivalent question, under a different guise, is posed in~\cite{fhns-hcag-00}:
Is the arrangement graph of great circles on the sphere $3$-colorable?
Assume no three circles meet at a point, so that this graph is $4$-regular.
Circle graphs in the plane can require four colors~\cite{k-4c4vp-90},
so the key property in this problem is that the circles must be great.
All arrangement graphs of up to $11$ great circles have been verified
to be $3$-colorable by Oswin Aichholzer (August, 2002).
See~\cite{w-mrfch-02} for more details.

\problem{Kissing Circle Representation}
        {Therese Biedl}{University of Waterloo}{biedl@math.uwaterloo.ca}

It is known that any planar graph $G$ can be represented by
``kissing circles'': an interior-disjoint collection of circles,
one circle per vertex, such that two circles touch (``kiss'') precisely when
the corresponding vertices are adjacent.
However, the computation of such kissing circles is not straightforward.
See \cite{k-kka-35, m-ptcpa-93, bs-rpg-93, s-cgpcm-94, s-accnc-91, z-lp-95-kissing}
for more information.

Suppose one loosens the kissing requirement and seeks instead a collection
of disks whose intersection graph is $G$.  Is it easier to compute
such a representation?
Can the disk centers be restricted to rational coordinates?
Can they be integers bounded by a polynomial in some parameters of the graph?

\problem{$3$-Manifolds Built of Boxes}
        {Joseph O'Rourke}{Smith College}{orourke@cs.smith.edu}

A result in~\cite{do-npbr-01,do-npbr-02} may be interpreted as
follows:
For any polyhedral $2$-manifold homeomorphic to a sphere 
$\mathbb{S}^2 \subset \mathbb{R}^3$,
all of whose facets are rectangles,
adjacent facets either meet orthogonally or are coplanar.
This raises the analogous question one dimension higher:
For any polyhedral $3$-manifold homeomorphic to a sphere
$\mathbb{S}^3 \subset \mathbb{R}^4$,
all of whose facets are rectangular boxes,
is it true that
adjacent facets lie either in orthogonal $3$-flats or within the same $3$-flat?
Very roughly, must a $3$-manifold built from boxes be itself orthogonal?

\problem{Visibility Product Characterization}
        {Tom Shermer}{Simon Fraser University}{shermer@cs.sfu.ca}

\newcommand\VP{{\rm VP}}
Let $P$ be a polygon, treated as a region in the plane.
Define (for lack of a better term) the
\emph{visibility product} $\VP(P)$ to be the following
four-dimensional set:
\[
\begin{array}{l}
  \VP(P) = \{ (x_1,y_1,x_2,y_2) \ |\ \\
  \phantom{\VP(P) = \{}
  (x_1,y_1) \in P, \; (x_2,y_2) \in P, \; \\
  \phantom{\VP(P) = \{}
  (x_1,y_1) \mathrm{\ can\ see\ } (x_2,y_2) \} 
\end{array}
\]
%Thus points $(x_1,y_1)$ and $(x_2,y_2)$ are visible to one another.
Two points can \emph{see} one another if the line segment between those points
is a subset of~$P$.
Thus $\VP$ is something like a set product capturing visibility.
Determine the structure of $\VP(P)$, characterize the set, 
find an algorithm to construct it, and determine if it has
utility.

\problem{3D Orthogonal Graph Drawings}
        {David Wood}{Carleton University}{davidw@scs.carleton.ca}
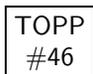
\TOPP{46}

Does every simple graph with maximum vertex degree $\Delta \leq 6$
have a 3D orthogonal point-drawing with no more than two bends per edge?
An \emph{orthogonal point-drawing} of a graph maps each vertex to a unique
point of the 3D cubic lattice, and maps each edge to a lattice path between
the endpoints; these paths can only intersect at common endpoints.
In this problem, each path must have at most two bends, that is,
consist of at most three orthogonal line segments (links).

There are several related known results.
Two bends would be best possible, because any drawing of $K_5$ uses
at least two bends on at least one edge.
If $\Delta \leq 5$, two bends per edge suffice \cite{w-3dogd-03}.
% above is also in \cite{w-a3dogd-98}
Two bends also suffice for the complete multipartite 6-regular graphs
$K_7$, $K_{2,2,2,2}$, $K_{3,3,3}$, and $K_{6,6}$
\cite{w-3dogd-00}.
In general, there is a drawing with an average number of bends per edge
of at most $2+\frac{2}{7}$ \cite{w-3dogd-03}.
% above has weaker result in \cite{w-a3dogd-98}
Additionally, three bends per edge always suffice,
even for multigraphs \cite{esw-tatdo-00, pt-aiogd-99, w-mnbv3d-01}.

% References and information thanks to
% ``Orthogonal Drawings With Few Layers'' by Biedl, Johansen, Shermer, Wood
% (GD'01)
% http://citeseer.nj.nec.com/biedl02orthogonal.html

This problem was first posed in \cite{esw-tatdo-00}.

\problem{Sailor-in-the-Fog Generalization}
        {Alejandro L\'opez-Ortiz}{University of Waterloo}{alopez-o@uwaterloo.ca}

The venerable ``Sailor in the Fog'' problem asks for an optimal
search strategy for a sailor to find the shoreline when
lost in a fog offshore (a version was posed by Bellman in~\cite{b-dp-87}).
There are many variations on this problem.
For example, one version can be rephrased as follows:
Find the shortest-length path from
the center of a unit disk that intersects every halfplane
whose bounding line (the shoreline) supports the disk.
Note here the assumption is that the distance to the shore
is known.
This problem was solved by Isbell \cite{i-osp-57}.

A conference conversation suggested the following higher-dimensional
generalization:
Find the shortest-length path from
the center of a unit ball that intersects every halfspace
whose bounding plane supports the ball.
This problem might represent a diver seeking the surface.

It came to light after the presentation that this problem
was posed before, in a paper by V. A. Zalgaller~\cite{z-sils-92},
for which there is apparently no published translation
from the Russian.
%which has apparently not been translated from the Russian.
Nonetheless, the problem remains unsolved.
See \cite{fw-lf-01} for more information.

\problem{Region Realization}
        {Alejandro L\'opez-Ortiz}{University of Waterloo}{alopez-o@uwaterloo.ca}

Suppose we are given a collection of constraints on unknown planar connected
regions of the form
\begin{enumerate}
\item region $A$ is contained in region $B$; and
\item region $A$ properly intersects region $B$;
\item regions $A$ and $B$ are adjacent (share just boundary points);
\item region $A$ does not touch region $B$.
\end{enumerate}
Is there a polynomial-time algorithm to decide whether there is a realization
of these constraints by planar connected regions?
The special case involving just constraints of Type 3 is called
a \emph{map graph}, a concept introduced by Chen, Grigni, and
Papadimitriou~\cite{cgp-mg-02} and solved by Mikkel Thorup \cite{t-mgpt-98}.

A simpler variation on this problem is that all regions are given (as planar
polygons) except for one unknown region $X$ which must be found in order to
obey the given constraints.

\problem{Guaranteed Aspect Ratio Partitions}
        {Mirela Damian-Iordache}{Villanova University}{mirela.damian@villanova.edu}

Define the \emph{aspect ratio} of a polygon as the ratio of
the diameters of the smallest circumscribing circle to the largest
inscribed circle.  (Thus in this context the aspect ratio measures
circularity.)
Find a polynomial-time algorithm for partitioning a 
polygon into the fewest polygonal pieces,
each piece with an aspect ratio no more than a given $\alpha > 1$,
or to report that no such partition exists.
Here the pieces are permitted to employ ``Steiner points,'' points
that are not vertices of the given polygon.
When Steiner points are disallowed, a polynomial-time algorithm
is known~\cite{d-eaaco-02}.
A second question is to find the smallest $\alpha > 1$ for which there
is a partition in which every piece has aspect ratio at most~$\alpha$.

\problem{Representing Separation by Pseudotriangulation}
        {Bettina Speckmann}{ETH Z\"urich}{speckman@inf.ethz.ch}

Consider a nonoverlapping collection of polygons in the plane.
Is it always possible to decompose the exterior of these polygons
into pseudotriangles such that each object touches at most
as many pseudotriangles as its minimum-link separation chain?
How efficiently can such a decomposition be computed, when it exists?

\problem{D-forms}
        {Joseph O'Rourke}{Smith College}{orourke@cs.smith.edu}

Let $c_1$ and $c_2$ be two smooth, closed, convex, planar curves of the same
length, each bounding a flat piece of paper.
Choose a point $p_1$ on $c_1$ and a point $p_2$ on $c_2$,
and glue the two curves to each other (according to arclength)
starting with $p_1$ glued to $p_2$.
The resulting single piece of paper forms a shape in space
called a \emph{D-form} by Helmut Pottmann, Johannes Wallner, and 
Tony Wills.
%\xxx[Erik]{Who is Tony Wills?  He's not an author of the book.}
The curves $c_1$ and $c_2$ join to form a closed space curve $\mathbf{c}$
bounding two developable surfaces $S_1$ and $S_2$.
These authors ask two questions in~\cite[p.~418]{pw-clg-00}:
\begin{enumerate}
\item
``It is not clear under what conditions a D-form is the convex
hull of a space curve.''
\item
``After some experiments we found that, surprisingly,
both $S_1$ and $S_2$ were free of creases, but we do not know whether
this will be so in all cases.''
\end{enumerate}

\end{description}

\end{document}